\documentclass[manuscript]{emulateapj}

\shorttitle{'No Smoking' zone for galaxies?}
\shortauthors{Khosroshahi et al.}

\begin{document}

\title{Galaxy And Mass Assembly (GAMA): 'No Smoking' zone for giant elliptical galaxies?}

\author{Habib G. Khosroshahi,\altaffilmark{1,2}
	Mojtaba Raouf,\altaffilmark{1}
	Halime Miraghaei,\altaffilmark{1}
	Sarah Brough,\altaffilmark{3} 
	Darren J. Croton,\altaffilmark{4} 
	Simon Driver,\altaffilmark{5} 
	Alister Graham,\altaffilmark{4} 
	Ivan Baldry,\altaffilmark{6} 
	Michael Brown,\altaffilmark{7}
	Matt Prescott,\altaffilmark{8}
	and Lingyu Wang\altaffilmark{9,10}}
\affil{$^1$School of Astronomy, Institute for Research in Fundamental Sciences (IPM), Tehran, 19395-5746, Iran}	
\affil{$^2$Institut d'Astrophysique de Paris, 98 bis Bd Arago, F-75014 Paris, France}	
\affil{$^3$School of Physics, University of New South Wales, NSW 2052, Australia}	
\affil{$^4$Centre for Astrophysics \& Supercomputing, Swinburne University of Technology, PO Box 218, Hawthorn, Victoria 3122, Australia}
\affil{$^5$International Centre for Radio Astronomy Research (ICRAR), The University of Western Australia, 35 Stirling Highway, Crawley, WA 6009, Australia}
\affil{$^6$Astrophysics Research Institute, Liverpool John Moores University, IC2, Liverpool Science Park, 146 Brownlow Hill, Liverpool L3 5RF, UK}
\affil{$^7$School of Physics, Monash University, Clayton, VIC 3800, Australia}
\affil{$^8$Astrophysics Group, The University of Western Cape, Robert Sobukwe Road, Bellville 7530, South Africa}
\affil{$^9$SRON Netherlands Institute for Space Research, Landleven 12, 9747 AD, Groningen, The Netherlands}
\affil{$^{10}$Kapteyn Astronomical Institute, University of Groningen, Postbus 800, 9700 AV, Groningen, The Netherlands}
\email{habib@ipm.ir}

%\altaffiltext{*}{Research Associate, IAP, Paris}
\begin{abstract}
We study the radio emission of the most massive galaxies in a sample of dynamically relaxed and un-relaxed galaxy groups from Galaxy and Mass Assembly (GAMA). The dynamical state of the group is defined by the stellar dominance of the brightest group galaxy, e.g. the luminosity gap between the two most luminous members, and the offset between the position of the brightest group galaxy and the luminosity centroid of the group. We find that the radio luminosity of the most massive galaxy in the group strongly depends on its environment, such that the brightest group galaxies in dynamically young (evolving) groups are an order of magnitude more luminous in the radio than those with a similar stellar mass but residing in dynamically old (relaxed) groups. This observation has been successfully reproduced by a newly developed semi-analytic model which allows us to explore the various causes of these findings. We find that the fraction of radio loud brightest group galaxies in the observed dynamically young groups is $\sim2$ times that in the dynamically old groups. We discuss the implications of this observational constraint on the central galaxy properties in the context of galaxy mergers and the super-massive blackhole accretion rate.

\end{abstract}


\section{Introduction}

Heating of the Inter-Galactic Medium (IGM) within the core of galaxy groups and clusters is an unresolved problem in extra-galactic astronomy \citep{Fabian2012, Mcnamara2007}. The gas in this region does not cool as dramatically as expected from the emitted X-ray emission \citep{David2001, Peterson2006}. Among the mechanisms proposed and discussed to balance the expected cooling, the role of Active Galactic Nuclei (AGN) feedback is seen as the most prominent \citep{Gitti2012, Blanton2010, Gaspari2011}, though the exact mechanisms are still debated. Mechanical heating \citep{Birzan04, Nulsen07, JHL14}, turbulence \citep{Zhurva2014}, mixing \citep{Gilkis2012}, deposition of energy through AGN originated shocks \citep{Graham2008, Randall2015} and sound waves \citep{Fabian2006} are among mechanisms through which the AGN could heat up the IGM in the group/cluster core. The evolution of galaxies and in particular the most massive galaxies in the Universe found in the core of groups and clusters is clearly affected by AGN feedback \citep{Mcnamara2007}, also widely assumed in galaxy formation and evolution models \citep{Croton2006}.

In order to understand whether galaxy environment influences AGN activity and thus feedback, we focus on dynamically relaxed galaxy groups also known as fossil groups \citep{Ponman1994,Khosroshahi2004,Khosroshahi2007}. The main characteristic of fossil groups is the stellar dominance of the Brightest Group Galaxy (BGG) generally probed by the optical luminosity or magnitude gap (e.g. $\Delta m_{12}\ge2.0$) within the half-virial radius or in a fixed projected radius of the group halo \citep{Jones2003}.  The conventional argument for the formation of fossil groups is based on a scenario in which a massive galaxy forms via cannibalizing its surrounding galaxies through dynamical friction, which requires several Gyr \citep{Jones2000}. A number of studies using cosmological simulations have shown that fossil galaxy group halos form relatively earlier than halos with a small luminosity gap (e.g. $\Delta m_{12}\le0.5$) and the results are consistent between hydrodynamical approaches \citep{DOnghia2005,Cui2011,Raouf2016a} and semi-analytical models for galaxies \citep{Sales2007,Dariush2007,Diaz2008,Dariush2010,Raouf2014}. The observational findings are seen to be consistent with the broad picture that the groups with a large luminosity gap have an earlier formation epoch than those with a small luminosity gap \citep{Khosroshahi2006a, Khosroshahi2007, Smith2010}. 

AGNs are powered by gas accretion on to the super-massive black hole at the center of galaxies and have been argued to significantly affect the evolution of the host galaxy through quenching of the star formation and also affect the IGM heating through various feedback processes. The accretion of hot gas (Bondi accretion) is tightly correlated to the AGN jet power \citep{Allen2006}. Cold accretion \citep{Werner2014} and black hole spin \citep{Russell2013} have also been explored to find out the main fueling mechanism. In a recent study of a small sample of fossil galaxy groups, we found indications that the most luminous galaxies in fossil galaxy groups, a representative for dynamically relaxed halos, are under-luminous in radio emission in 610 MHz and 1.4 GHz  \citep{Miraghaei2014}. 
This study suggests that mergers, the key phenomena behind the formation of a  large luminosity gap, may be the main source of discrimination in the radio properties. In a more recent study we have examined the IGM heating sources in one of the most massive fossil groups \citep{Miraghaei2015} and found that in the case of RX J1416.4+2315, the energy injected into the IGM by AGN is only sufficient to heat up the central 50 kpc.
%\clearpage

\begin{figure}
	%\epsscale{.80}
	%\plotone{redshift_distribution.pdf}
	\includegraphics[width=0.47\textwidth]{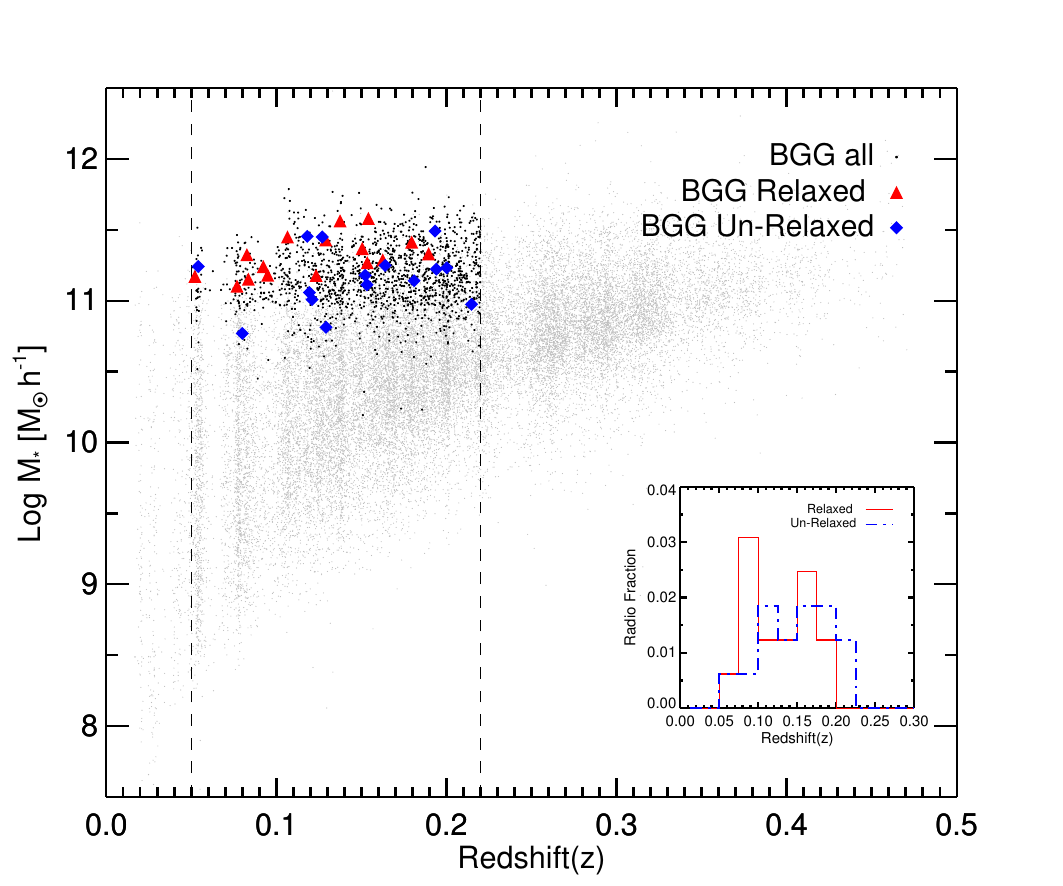}
	\caption{Selection function for the samples; the stellar mass of the BGGs as a function of the redshift. The background grey dots represent all galaxies assigned to groups in the entire GAMA database. The black dots represent luminous BGGs ($M_r\leq-22$) within the redshift limit of the sample which is defined based on the redshift completeness of the sample. The symbols represent BGGs in dynamically relaxed and un-relaxed groups with a radio detection in 1.4 GHz with the sub-panel representing their redshift distribution. \label{fig1}}
\end{figure}

%\clearpage
\begin{figure*}
	\centering
	%\plotone{radio_stellarmass_fossil_ctrl_STD_peak_int.pdf}
	\includegraphics[width=0.9\textwidth]{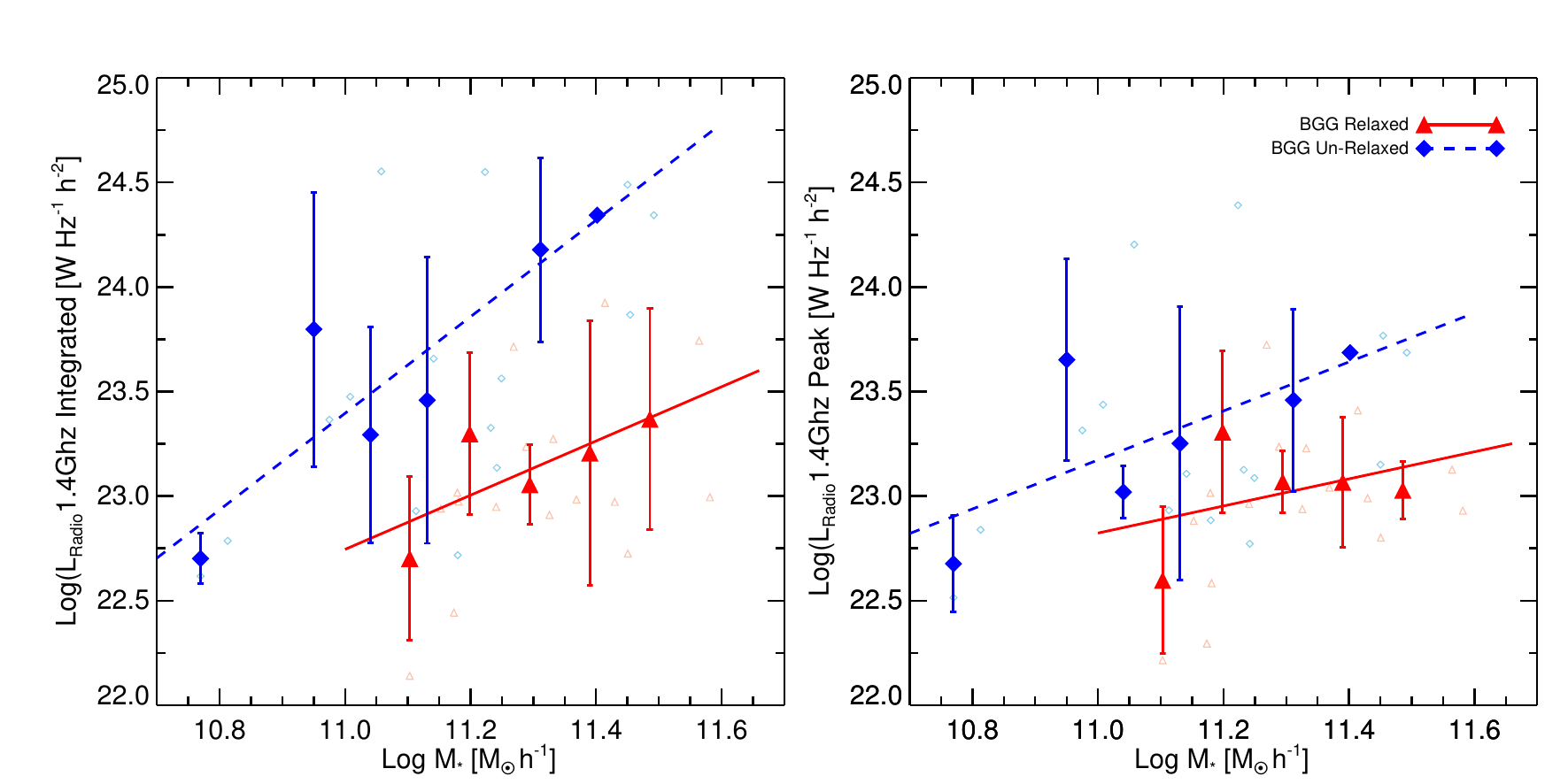}
	\caption{The 1.4 GHz radio power of the brightest group galaxy in old or dynamically relaxed groups (red) and young or evolving groups (blue). Old galaxy groups are selected to have a large luminosity gap ($\Delta m_{12}\ge1.7$) and a relatively small off-set between the BGG and the luminosity centroid of the group ($\le$100 kpc). Young groups have small luminosity gap ($\Delta m_{12}\le0.3$) and a large off-set ($\ge$100 kpc). The radio luminosity refers to the integrated (left) and peak (right) flux densities obtained from the VLA First catalog. The symbols mark the median value over the bin (0.1 in log scale) and the small symbols represent individual BGGs. The scatter in also shown for the binned data. The solid/dashed lines indicate linear regressions to the binned data.\label{fig2}}
\end{figure*}

A number of  studies found an increased fraction of AGNs in galaxies with close matches, ongoing mergers or in post-merger systems \citep{Almeida2010,Ellison2011,Bessiere2012,Cotini2013,Sabater2013}. However, several studies claimed to find no significant excess of mergers in AGN hosts \citep{Dunlop2003,Sanchez2004,Grogin2005,Li2008,Gabor2009,Tal2009,Cisternas2011,Kocevski2012,Schawinski2012,Bohm2013}. These studies report that a significant fraction of the AGN appear to reside in isolated disc-dominated galaxies for which internal processes are likely responsible for fueling their active nuclei. The focus of our study is the brightest galaxy in the group which are primarily giant elliptical galaxies. Assuming a simple picture for the gravitational collapse of gas into the center of galaxy cluster, BCGs located at the cluster center will be influenced by a larger reservoir (density) of hot gas compared to BCGs with a large offset. 

Galaxy groups in different dynamical states are suitable systems to study the AGN fueling and its feedback on galaxy evolution, because of the absence of recent group scale mergers and galaxy major mergers in virilized groups compared to evolving groups \citep{Jones2003}. Although the luminosity gap and the offset between the BGG and the luminosity centroid are both indicators for the dynamical state/age of galaxy groups \citep{Raouf2014}, the luminosity gap is a key player.  The presence of a large luminosity gap points at the absence of recent major merger which could ignite the cold mode accretion. While an AGN at the bottom of the potential well of the group/cluster, where the IGM reaches its peak density in a dynamically relaxed group is subject to hot gas accretion. In Section 2 of this paper we describe the sample and the data. Section 3 is dedicated to the analysis followed by a discussion and concluding remarks in section 4.

Throughout this paper  we use a $\Lambda CDM$ cosmology with $\Omega_m=0.3$, $\Omega_\Lambda=0.7$ and $H_0=70 $ km s$^{-1}$ Mpc$^{-1}$.

\section{Data and sample selection}
The main source of data for this study is the Galaxy And Mass Assembly (GAMA) survey, a  
multi-wavelength spectroscopic data set covering an area of 180 deg$^2$. The description of the survey is given in \cite{Baldry2010} while other aspects of the survey have been described in \cite{Robotham2010}, \cite{Driver2011} and  \cite{Hopkins2013}. We use the second data release, GAMA-DRII.

%, which includes the GAMA galaxy group catalog  \citep{Robotham2011}.
We use GAMA stellar masses catalog which provides stellar masses, rest frame photometry, and other ancillary stellar population parameters from stellar population fits to {\em "ugriz"} Spectral Energy Distributions (SEDs) for all $z < 0.65$ galaxies \citep{Taylor2011}. 

The GAMA-DRII galaxy group catalog has been generated using a friends-of-friends (FoF) based grouping algorithm \cite{Robotham2011}. The catalog contains 23,838 galaxy groups which reduces to about 2,500 galaxy groups and about 19,000 group members with multiplicity of at least 4 spectroscopically confirmed members.  Using the total extrapolate luminosity and the total stellar mass of the group galaxies and their positions we obtain the luminosity gap and the luminosity centroid of the groups. 
We select a sample of dynamically relaxed galaxy groups and a control sample (dynamically evolving groups) as follows:

I. A sample of galaxy groups with a brightest group galaxy at least as luminous as $M_r=-22$ mag (total of 1533 groups) and with a large luminosity gap between the BGG and the second brightest group member, $\Delta$m$_{12}\ge1.7$ in r-band. In addition we also impose that the BGG is located within a radius of 100 kpc of the luminosity/stellar-mass centroid of the group. This results in 174 groups. 

II. A sample of galaxy groups with a BGG to at least as luminous as $M_r=-22$ mag and with a small luminosity gap, $\Delta$m$_{12}\le 0.3$ in r-band. We impose the BGG to be located outside the radius of 100 kpc centered on the luminosity/stellar-mass centroid of the group. This results in 134 groups. We note that the majority of galaxy groups tend to have a small luminosity gap \citep{Gozali14}, however the large off-set requirement reduces the sample to a size comparable to Sample I.

The luminosity centroid of the group members is provided in the GAMA group catalog and is defined as the center of light derived from the r-band luminosity of all the galaxies identified to be within the group \citep{Robotham2011}. The redsift limit is chosen on the basis of providing a complete sample of groups with a luminosity gap of $1.7$ mag. We cross match the brightest group galaxies with the VLA FIRST catalog of objects detected in 1.4 GHz. The FIRST survey has released a catalog which contains all the radio sources detected above a limiting flux density of $\sim$ 1 mJy for point sources with a typical noise of $\sigma$ $\sim$ 0.13 mJy \citep{Becker1995}. About 10\% of the luminous BGGs ($M_r \leq -22$) in the sample are associated with a FIRST catalog source. Among the radio detected BGGs, about 10\% are assigned to relaxed groups and an equal fraction are assigned to un-relaxed groups, with a redshift distribution shown in the subpanel in Fig 1.  The focus of the study will be on the BGGs hosted by relaxed and unrelxed groups with a detected radio emission. The low frequency radio emission at 325 MHz has been obtained for the GAMA fields using GMRT observations  \citep{Mauch2013} with 14-24 arcsec resolution and $\sim$ 10 mJy limiting flux density. The low frequency observations allow us to study low energy electrons and thus the past AGN activities of galaxies \citep{Miraghaei2014}. 

Figure \ref{fig1} shows the selection function of the sample highlights the stellar mass of the brightest group galaxies and the redshift distribution of the groups (e.g. the redshift associated with the brightest group galaxy). The small difference between the adopted $\Delta m_{12}=1.7$ limit used for the selection of the relaxed groups and the one conventionally used in previous studies of fossil groups, $\Delta m_{12}=2.0$, is to ensure a statistically meaningful number of galaxies in both the above samples. Other authors have also adapted similar variations in the sample selection of fossil galaxy groups \citep[e.g.,][]{Gozali14}. 

\section{Analysis and results}

We first attempt to establish whether the ongoing AGN activity in the BGGs probed by the 1.4 GHz luminosity is influenced by the dynamical state of the group, based on the two aforementioned halo age indicators, the luminosity gap and the BGG offset. The observed correlations between the masses of black holes in the nuclei of nearby galaxies and global galactic properties, as the bulge luminosity or the central velocity dispersion, point towards a direct link between the physical processes that contribute to the central black hole's growth and the formation of their host galaxies.

In Figure \ref{fig2} we present the distribution of the BGG radio emission as a function of the galaxy stellar mass for the two samples. We quantify the relation between the radio luminosity and the stellar mass of the brightest group galaxy using a linear regression, $log(L_{radio}) =a (log(M_*)) + b$, where $a$ and $b$ are given in table \ref{TB:Slope} for different samples. 
\begin{table}
	\centering
	\scriptsize
	\caption{The slope and intercept ($a$ and $b$) of the radio luminosity -- stellar mass relation for our samples. $^{(*)}$ Note the limited statistics. $^{(\dag)}$ Marks upper limit.}
	\label{TB:Slope}
	\begin{tabular}{lccc}
		\hline\hline
		Group & $a$ & $b$ & Count \\
		\hline\hline
		Relaxed (1.4 GHz integrated) & 1.47$\pm$0.76 &6.55$\pm$8.54 & 16 \\
		UnRelaxed (1.4 GHz integrated) & 2.45$\pm$0.75&-3.65$\pm$8.38 & 15\\	
		Relaxed (1.4 GHz peak) & 0.65$\pm$0.91 &15.69$\pm$10.26 & 16\\
		UnRelaxed (1.4 GHz peak) & 1.17$\pm$0.6&10.28$\pm$6.68 & 15\\	
		Relaxed (325 MHz) & 1.71$\pm$0.0$^*$ &4.29$\pm$0.0$^*$ & 13$^\dag$\\
		UnRelaxed (325 MHz) & 0.7$\pm$0.95&16.52$\pm$10.7 & 13\\	
	    Fossil ($\Delta m_{12}>1.7$) & 	1.08$\pm$0.21&10.87$\pm$2.39 & 29\\
		Non-Fossil ($\Delta m_{12}<0.3$) &  0.46$\pm$0.53&18.08$\pm$5.88 & 23\\
		Low off-set ($D_{off-centr}<100 kpc$ )& 	0.6$\pm$0.36&16.65$\pm$3.97 & 65\\
		High off-set ($D_{off-centr}>100 kpc$) &  0.98$\pm$0.30&12.26$\pm$3.34 & 73\\
		Old (SAM) & 3.2$\pm$0.08&-12.97$\pm$0.86 & 2515\\
		Young (SAM) & 2.43$\pm$0.11&-3.63$\pm$1.23 & 458\\
		\hline\hline
	\end{tabular}
	
\end{table}
This is a clear demonstration that the relaxed (old) galaxy groups harbor BGGs that are less radio luminous in comparison to BGGs in groups which are classified as un-relaxed or  evolving (young). The difference in the radio luminosity of the BGGs in these two dynamically different environments is measured to be a striking one order of magnitude in the radio luminosity, pointing to a significant, if not determining, influence of the environment on the AGN activities in the brightest group galaxies. The results are the same when we adapt both the peak and the integrated radio luminosity at the location of the BGGs. 

We argue that this is not an observational bias. For instance the source confusion may be seen as a possible reason for the observed difference. If radio luminosity of two or more galaxies, in the sample with the least luminosity gap, are attributed to the BGG due to poor angular resolution, the BGGs in the sample of evolving groups will appear over-luminous in the radio. The spatial resolution of the VLA First is $\le5$ arcsec. For the most distant group sample at a redshift of $\approx$0.3 , such an angular resolution corresponds to a physical size of $\approx 20$kpc. To eliminate any such source confusion bias, both samples are required to have at least a 60 kpc projected separation between the two most luminous galaxies. It is clear that, given the definition of the samples described above, this additional criteria will only affect the statistics in Sample II. However this is a small effect and finally the sample I and II contain 16 and 15 BGGs, respectively.  Furthermore, we use both the peak radio luminosity and the integrated radio luminosity in our comparisons to rule out such a bias. The visual inspection shows no indication of source confusion. This is an additional constraint on samples I and II described in section 2 with the final statistics presented in Table \ref{TB:Slope}. We remind that this 60 kpc cut between two most luminous galaxies in groups is different from the 100 kpc offset between the BGG and the group luminosity centroid.

\subsection{Luminosity gap vs off-centering }\label{comp}

While we have established that the BGGs in relaxed galaxy groups are strikingly under luminous in radio continuum emission, in comparison to those hosted by evolving groups, we now explore the origin of the observed difference. In particular we attempt to discriminate between the effect of galaxy mergers which is the driving phenomena behind the formation of the large luminosity gap and the role of the hot mode accretion which may be occurring given the privileged position of the BGG at the bottom of the potential well, where gas accretion is expected to be directed towards. We recognize that X-ray observations do not feature in our study, however, a recent study by \cite{Khosroshahi2014} has shown that under a similar selection criteria employed in this study, an extended X-ray emission, associated with the group halo, surrounds the giant elliptical galaxy.  

\begin{figure}
	\includegraphics[width=0.47\textwidth]{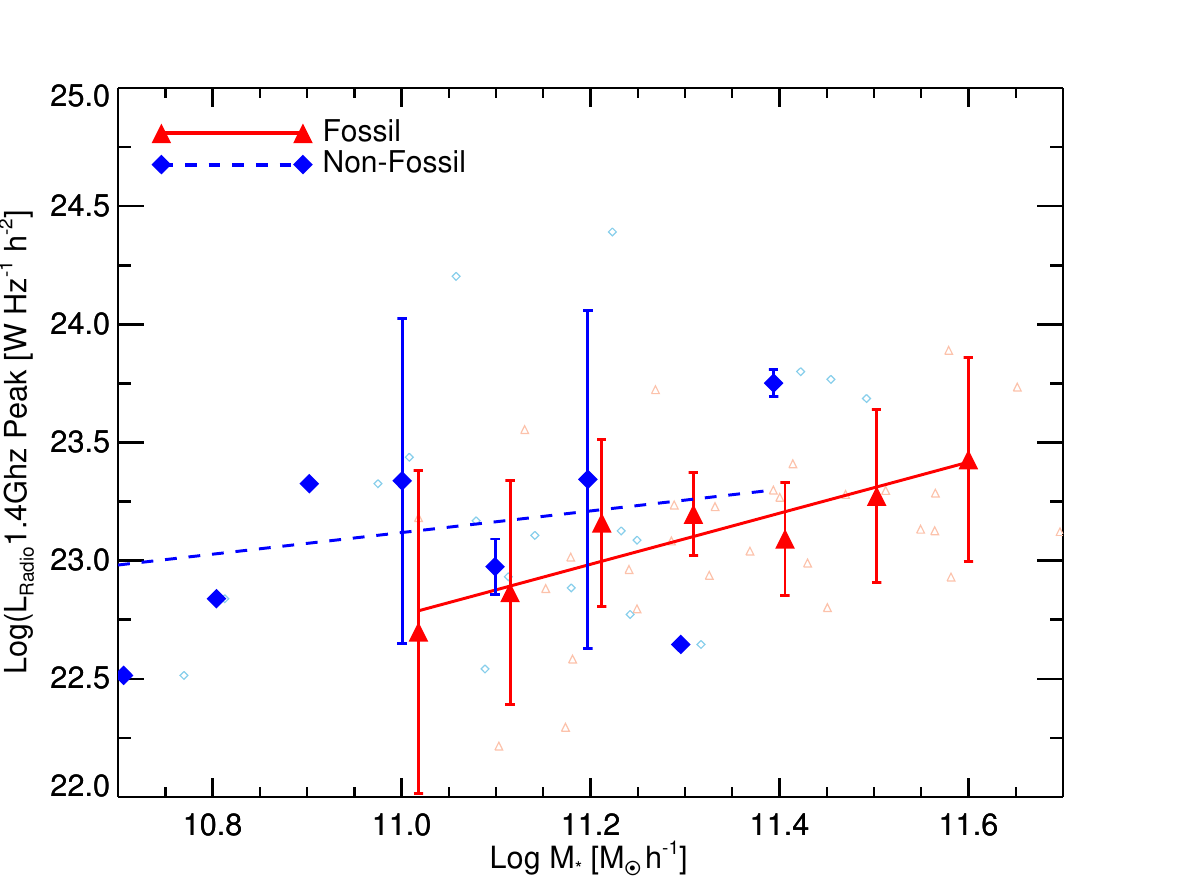}
	\caption{The 1.4 GHz radio power of the brightest group galaxy in Fossil groups ($\Delta m_{12}\ge1.7$, red) and Non-Fossil group ($\Delta m_{12}\le0.3$, blue). The radio luminosity refers to the peak flux density obtained from the VLA First catalog. The bold and small size symbols refer to the average bin and the individual BGGs, respectively.\label{fig3}}
\end{figure}

We thus relax the offset criterion for the BGG and only keep the constraint on the luminosity gap in both samples. The difference between the radio luminosity of the BGGs in large and small luminosity gap groups is significantly reduced and thus the BGGs in large luminosity gap systems, conventionally known as fossil groups, are marginally less luminous in the radio (Figure \ref{fig3}) than in the control sample in which the luminosity gap is very small.  Thus the cold mode accretion due to mergers does not appear to the driving phenomena behind the observed difference in the radio luminosity of the BGGs in the two samples and as a result their AGN activities.

\begin{figure}
	\includegraphics[width=0.47\textwidth]{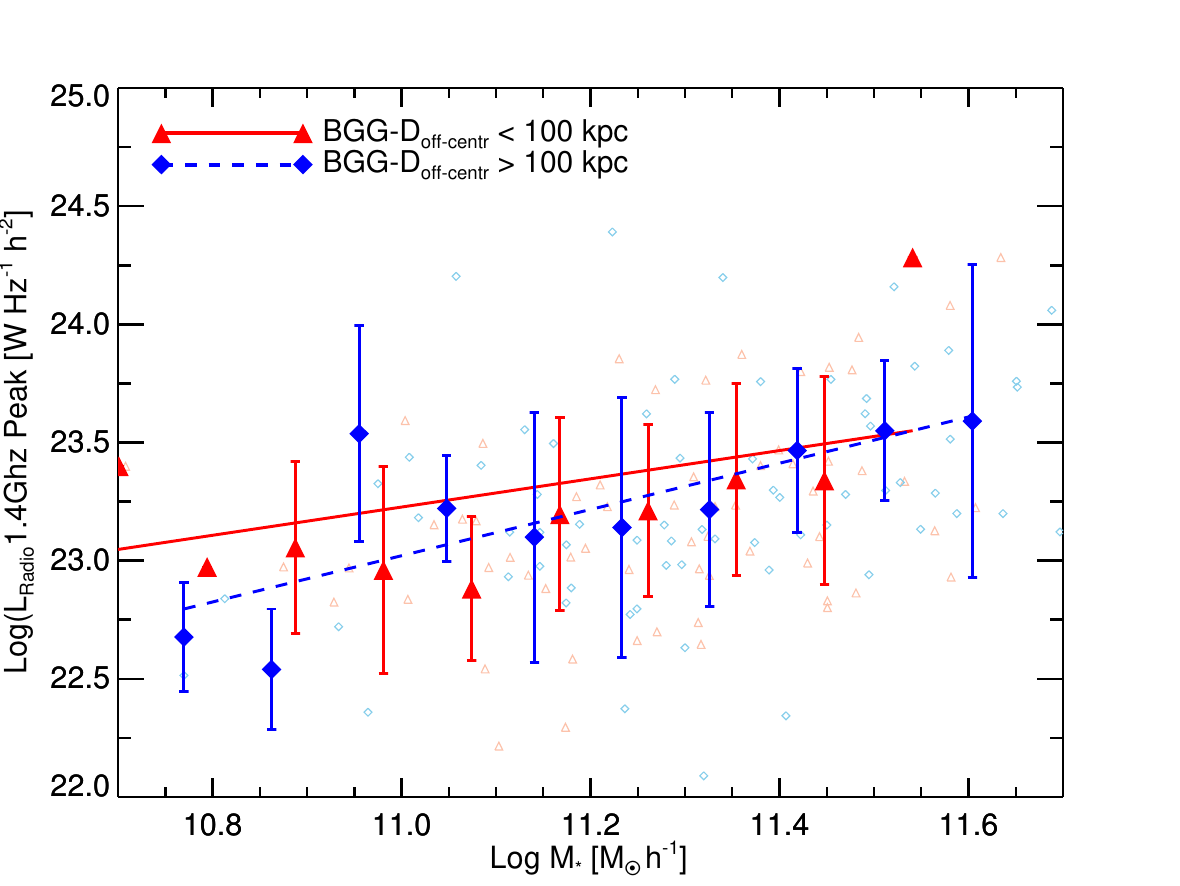}
	\caption{The 1.4 GHz radio power of the brightest group galaxy in groups with a small off-centring between the BGG position and the luminosity centroid of the group ($\le$100 kpc, red) and large off-center groups ($\ge$100 kpc, blue). The radio luminosity refers to the peak flux density obtained from the VLA First catalog. No significant difference between the radio power of the BGGs is seen in two samples.  The bold and small size symbols refer to the average bin and the individual BGGs, respectively. \label{fig4}}
\end{figure}

We relax the luminosity gap criterion to study the role of the offset between the BGG and the luminosity centroid of the galaxies in order to find out if a large centroid offset, which can disrupt the hot mode accretion on to the central galaxy, plays a role in the AGN activity. Figure \ref{fig4} shows that such an offset does not influence the BGG radio luminosity. We note that both the small number statistics and the absence of X-ray data, are two limiting factors in making a concrete statement on the influence of the BGG position within the group and its radio activity. It worth noting that \citet{Sanderson09} in a study of a sample of galaxy clusters found that the BCGs which are located within 15 kpc of the peak of the X-ray emission, are likely to be associated with the radio and line  emission, than those which show a larger centroid offset from the cluster core in an apparent contradiction to our finding in Figure \ref{fig4}, however the offset scale used in this study is larger.
     
\begin{table*}
\centering
\caption{Sample of brightest group galaxies in dynamically relaxed galaxy groups. The upper limit of 325 MHz luminosity for undetected sources are marked with $\dagger$ sign. Note that the group's halo mass are reported by GAMA group catalog \citep[see, ][]{Robotham2011}.  $^{(*)}$ Do not report by GAMA.}  
\label{Relaxe:table}
\begin{tabular}{ccccccccc}
\hline\hline
 GroupID & Ra   &     Dec   &   $\Delta m_{12}$  &  $D_{off-centr}$ &  $Log (L_{int})_{1.4 GHz}$ &  $Log  (L_{peak})_{1.4 GHz}$ & $Log  (L)_{325 MHz}$ & $Log (M_{halo})$ \\ 
   GAMA-ID   &   [deg]   &   [deg]   &    [mag]      &     [kpc]   &  [$W Hz^{-1} h^{-2}$] &  [$W Hz^{-1} h^{-2}$]& [$W Hz^{-1} h^{-2}$] & [$M_{\odot}$]\\
\hline\hline
  300395 & 214.64665 &-0.68262 &2.5  &52.5   &  22.98 &23.04 & 23.79$^{\dagger}$  &  12.75  \\
  200594 & 179.98355 &-0.32312 &2.2  &24.5   &  22.94 &22.88 & 23.24$^{\dagger}$&  13.97  \\
  300260 & 214.45775 &0.51094  &1.8  &78.2   &  22.44 &22.30 & no detection       &  14.66  \\
  300422 & 216.4731  &0.5803   &1.7   &77.9    &  22.97 &22.99 & 23.64$^{\dagger}$& 13.97   \\
  200204 & 176.83746 &-1.70279 &2.8  &20.7    &  22.95 &22.96 & 23.52 &   14.80 \\
  200122 & 176.79744 &-1.88907 &1.8  &52.6   &  22.97 &22.58 & 22.81$^{\dagger}$& 14.28   \\
  100172 & 140.3522  &-0.40958 &1.9   &58.3   &  22.99 &22.93 & 23.81$^{\dagger}$&  14.62  \\
  300083 & 219.16068 &1.18301  &3.0  &91.2   &  23.74 &23.13 & 24.07 &  13.90  \\
  300282 & 219.69284 &1.11603  &2.4  &61.2   &  22.91 &22.94 & 23.22$^{\dagger}$&  15.68  \\
  200524 & 178.85081 &1.96747  &2.2  &66.1   &  23.92 &23.41 & no detection       &  14.02  \\
  200873 & 180.59833 &1.87271  &1.9  &46.4   &  23.24 &23.24 & no detection       &  13.34  \\
  300874 & 216.37852 &1.93393  &1.9  &50.3   &  23.71 &23.72 & 23.81$^{\dagger}$&  --$^*$  \\
  300542 & 215.39764 &1.38464  &1.7  &85.6   &  23.27 &23.23 & 24.01$^{\dagger}$&  13.72  \\
  300390 & 213.25523 &-1.1241  &2.5  &29.7    &  22.14 &22.22 & 23.16$^{\dagger}$&  13.40  \\
  301140 & 217.60904 &-1.13174 &1.8  &57.3   &  23.02 &23.02 & 23.60$^{\dagger}$&   14.13 \\
  200025 & 175.44472 &-0.51805 &2.3  &84.5   &  22.73 &22.80 & 23.46$^{\dagger}$& 13.86   \\
  \hline\hline
\end{tabular}

\end{table*}

\begin{table*}
\centering
\caption{Sample of brightest group galaxies in dynamically un-relaxed (evolving) galaxy groups.  The upper limit of 325 MHz luminosity for undetected sources are marked with $\dagger$ sign.}
\label{unRelaxe:table}
\begin{tabular}{ccccccccc}
\hline\hline
 GroupID & Ra   &     Dec   &   $\Delta m_{12}$  &  $D_{off-centr}$ &  $Log (L_{int})_{1.4 GHz}$ &  $Log  (L_{peak})_{1.4 GHz}$ & $Log  (L)_{325 MHz}$ & $Log (M_{halo})$\\ 
   GAMA-ID   &   [deg]   &   [deg]   &    [mag]      &     [kpc]   &  [$W Hz^{-1} h^{-2}$] &  [$W Hz^{-1} h^{-2}$]& [$W Hz^{-1} h^{-2}$] & [$M_{\odot}$]\\
\hline\hline
  100046 & 140.65239  & -0.40903  & 0.2   & 159.4   &   23.13  & 22.77 & 23.45 &   13.88 \\
  200565 & 178.36423  & -1.18102  & 0.2   & 144.0   &   22.62  & 22.51 & 24.02 &  12.22  \\
  200043 & 184.70724  & -1.04693  & 0.1   & 153.7   &   23.87  & 23.77 & 24.29 &  14.69  \\
  300170 & 222.57996  & -1.11318  & 0.1    & 280.9   &   24.55  & 24.20 & no detection      &  13.69  \\
  300102 & 213.46605  & -0.6308   & 0.3    & 149.6   &   23.47  & 23.44 & 23.58$^{\dagger}$& 13.85   \\
  300033 & 213.73576  & 0.20641   & 0.1   & 119.4   &   24.48  & 23.15 & 24.74 &   14.17 \\
  301202 & 216.47969  & -0.27155  & 0.2   & 202.2   &   22.78   & 22.84 & 23.64$^{\dagger}$&  13.59  \\
  200435 & 185.30832  & -1.37898  & 0.0   & 101.6   &   22.72  & 22.88 & 23.80$^{\dagger}$&   13.84 \\
  200045 & 180.76495  & -1.93058  & 0.1   & 200.8   &   22.93  & 22.93 & 23.80$^{\dagger}$&  13.97  \\
  100079 & 137.97845  & 1.14878   & 0.1   & 267.6   &   23.56  & 23.09 & 24.76 &   14.32 \\
  300377 & 213.06422  & -1.13354  & 0.0   & 356.8   &   23.66  & 23.11 & 23.96 &  13.53  \\
  200022 & 183.08499  & 1.80787   & 0.1   & 272.3   &   24.34  & 23.68 & no detection      &  14.18  \\
  300392 & 213.03886  & -0.83542  & 0.2   & 159.2   &   24.55  & 24.39 & 24.90 &   13.19 \\
  100286 & 131.20869  & 1.60518   & 0.1   & 168.2   &   23.33  & 23.13 & 24.46 &   13.59 \\
  301381 & 214.6561   & 1.65797   & 0.1   & 149.6   &   23.37  & 23.31 & 24.13$^{\dagger}$&  12.84  \\
    \hline\hline
\end{tabular}
\end{table*} 

\subsection{The radio map }\label{map}
We visually inspected the radio map of these galaxies in the two categories. Roughly, 5 per cent of both samples have radio emission above 3 $\sigma$ of the maps shown in Figures  \ref{relaxed:fig} and \ref{Un_relaxed:fig}. The radio contours in 1.4 GHz from the VLA survey \citep{Becker1995} are overlaid on the optical images of the groups from Sloan Digital Sky Survey data release 12 \citep[SDSS-DR12]{Alam2015} archive.  

The radio contours of young evolving groups show the existence of both extended and point source emission while nearly all BGGs dominating relaxed galaxy groups show no extended emission. Single or double radio lobes have been detected in about 30 per cent of the un-relaxed group sample compared to 10 per cent for relaxed samples. The fraction of BGGs with radio loud AGNs (L$_{radio}\ge 10^{23}$ W Hz$^{-1}$) is 73 per cent in young systems while it is only 37 per cent in relaxed groups. 

\begin{figure}
	%\plotone{radio_325MHZ.pdf}
	\includegraphics[width=0.47\textwidth]{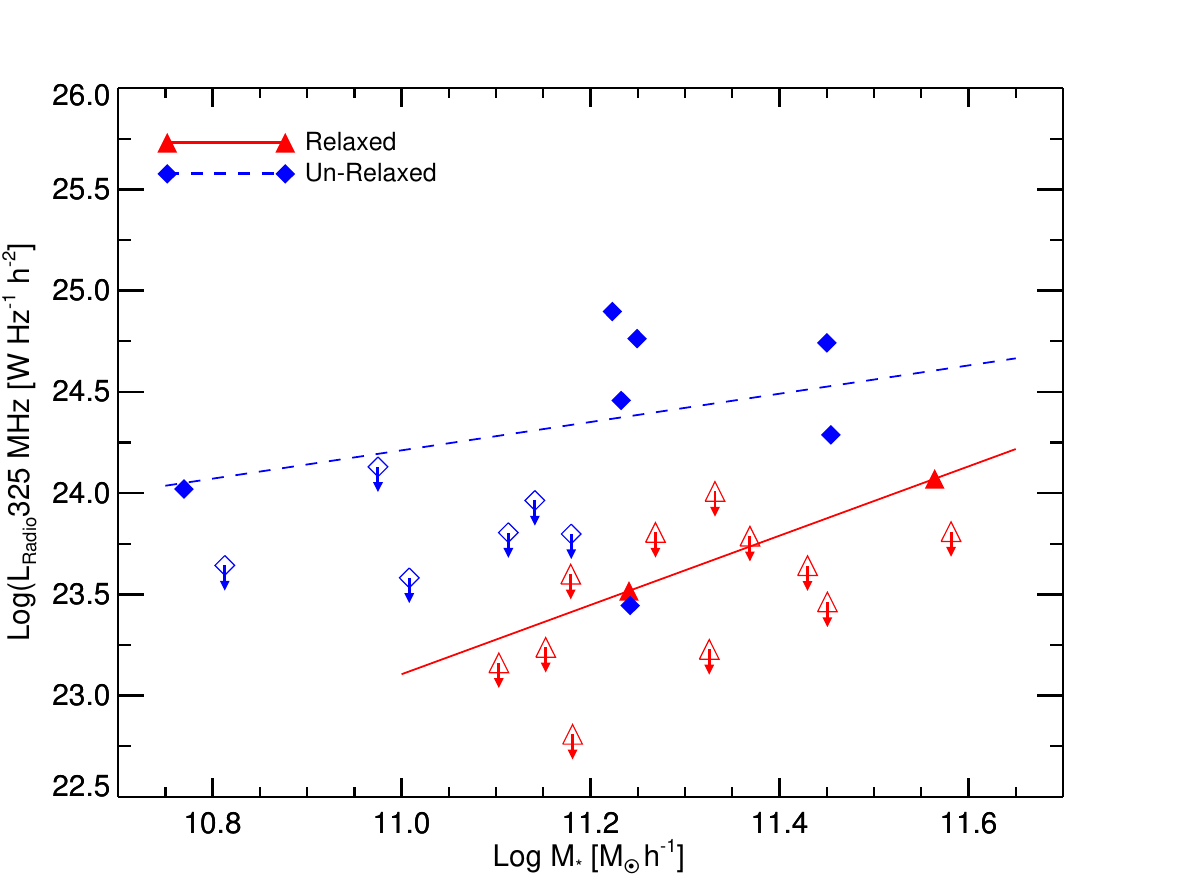}
	\caption{The 325 MHz radio power of the brightest group galaxies in relaxed (filled red) and un-relaxed (filled blue) groups. An  upper limit is given for the undetected relaxed (open red) and un-relaxed (open blue) BGGs. Majority of the BGGs in dynamically groups are radio loud.\label{fig5}}
\end{figure}

Furthermore, we adapted the 325 MHz GMRT radio luminosity to investigate the low frequency radio emission of the BGGs in relaxed and un-relaxed samples and the results are presented in Fig. \ref{fig5}.  The young (filled blue) and old (filled red) samples have been cross-matched with the 325 MHz GMRT observations of the GAMA field \citep{Mauch2013} within a 1 arcmin search radius. Given the sensitivity of the 325 MHz map, fewer number of objects have been detected in 325 MHz consisting of 2 and 7 objects in the old and young samples, respectively. Thus we also present an upper limit luminosity for the undetected objects in 325 MHz in Fig. \ref{fig5} (open markers) in which we used the 10 mJy flux density limit to calculate the luminosities. This further supports the results based on the 1.4 GHz measurements  where the BGGs in old groups are less luminous compared to those in the young evolving galaxy groups.  As the plot indicates, we find a higher fraction of radio loud sources among the BGGs in young evolving groups than in the old and relaxed galaxy groups. Only 7 percent of the BGGs in old systems show radio emission above L$_{325 MHz}\ge 10^{24}$ W Hz$^{-1}$ while 46 percent of BGGs in young groups have emission above this luminosity.

We limit our analysis to BGGs at least as bright as $M_{R} < -22$ mag to avoid late-type modest galaxies,  i.e. targeting the giant galaxies in both samples of relaxed and un-relaxed. However, two of the BGGs in the relaxed galaxy groups and one BGG in the un-relaxed galaxy groups are morphologically classified as spirals.

\subsection{Semi-analytic prediction}
The observational results described above are highly significant for understanding the AGN properties and its impact on galaxy evolution and environment dependent feedback. Given this, our efforts were focused on modelling the radio AGN in a cosmological context using the SAGE semi-analytic galaxy model \citep{Croton2016} and Millennium simulation\citep{Springel2005}. We developed a new method in which we trace the physical properties of radio jets in massive galaxies, including the evolution of radio lobes and their impact on the surrounding gas. In our model, we self-consistently trace the cooling-heating cycle that significantly shapes the life and death of many types of galaxies \citep[Radio-SAGE,][]{Raouf2017}. As the development of this model was motivated by the observations described above, the radio luminosity as an observable quantity, is calculated to allow us to study the effect of environment on AGN radio luminosity, radio luminosity function and properties of jet power in formation of host galaxies.  

\begin{figure*}
	\centering
	\includegraphics[width=0.9\textwidth]{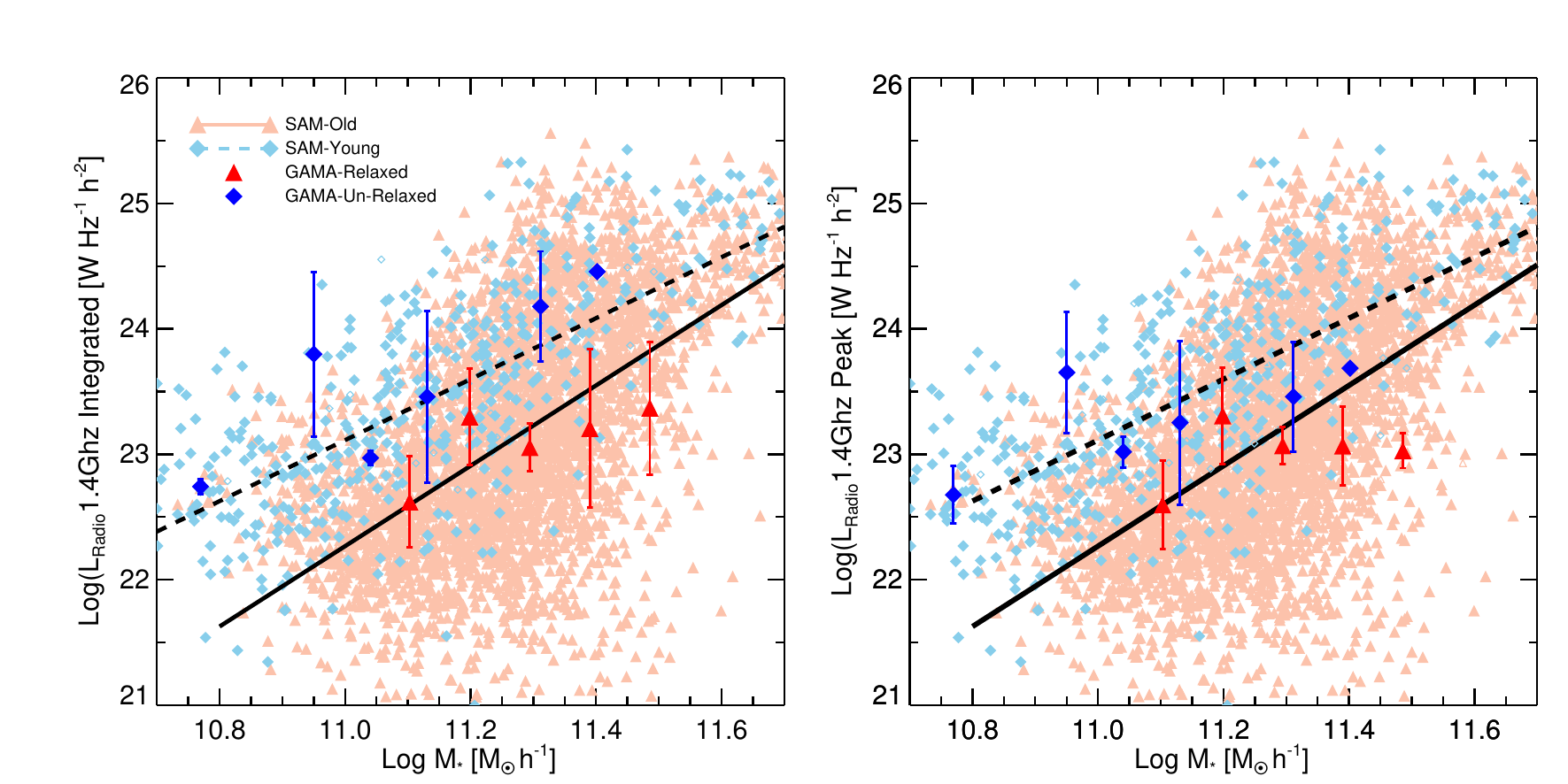}
	\caption{The 1.4 GHz radio power of the brightest group galaxy in old or dynamically relaxed groups (red) and young or evolving groups (blue) the same as in Fig. \ref{fig2}. The radio luminosity refers to the integrated (left) and peak (right) flux densities obtained from the VLA First catalog. The bold sized symbols indicate the average value over the bin. We overlay the central galaxies in old (light red triangles) and young (sky blue diamonds) galaxy groups corresponding to different stellar masses as function of radio luminosity predicted by our galaxy formations model. The solid and dashed-lines represent linear fits to the model data points for the BGGs in old and young galaxy groups, respectively.\label{fig6}}
\end{figure*}

\begin{figure*}
	\includegraphics[width=0.99\textwidth]{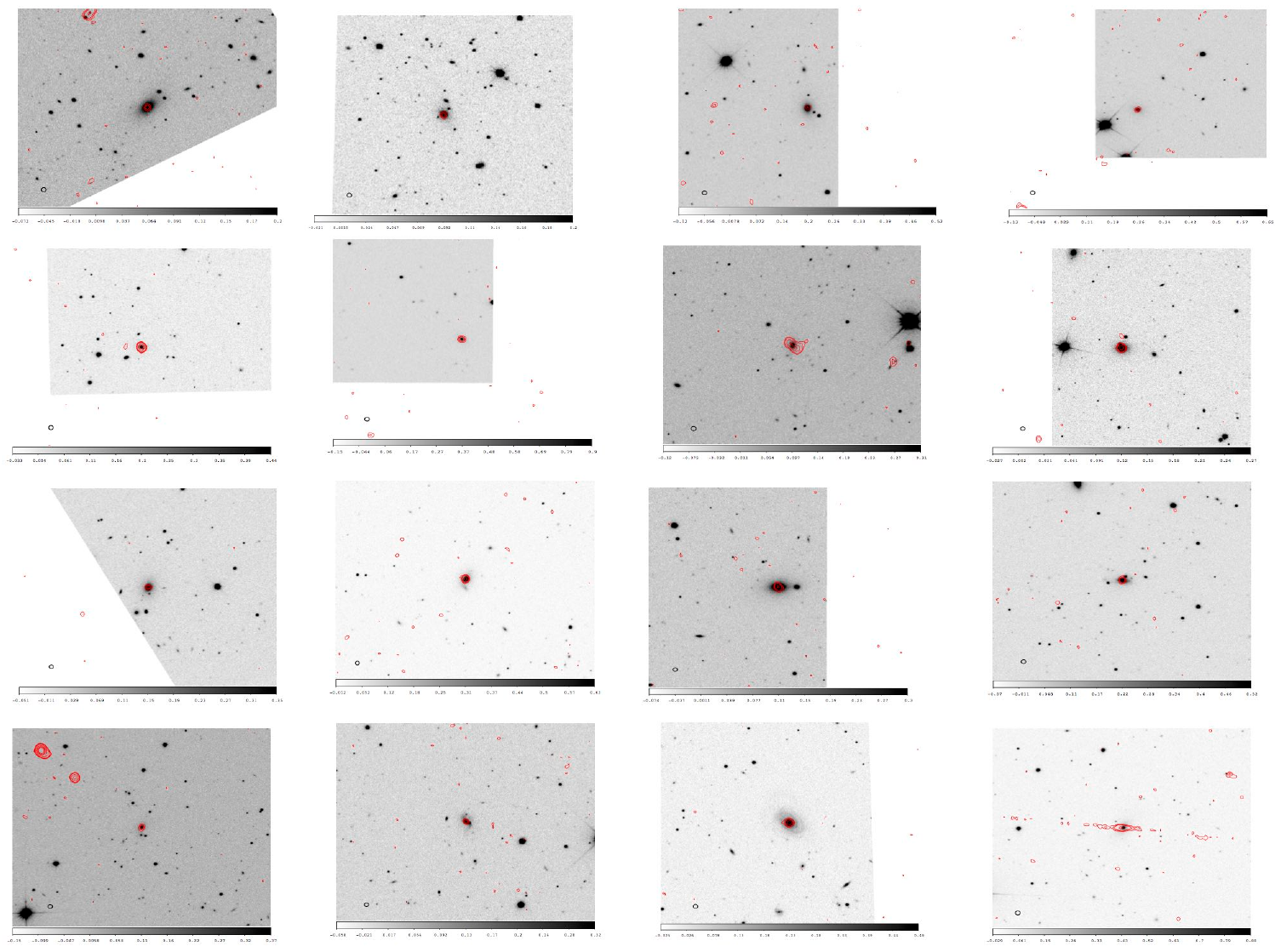}
	\caption{The radio contours overlaid on SDSS r-band image for relaxed (red) groups. Contour levels of 3, 5, 7, 19, 40, 80, 120, 180, 300, 500, 800, 1000 ${\sigma}$ are shown. The \textit{rms} noise ranges from 0.1-0.2 mJy in sample.}
	\label{relaxed:fig}
\end{figure*}

\begin{figure*}
	\includegraphics[width=0.99\textwidth]{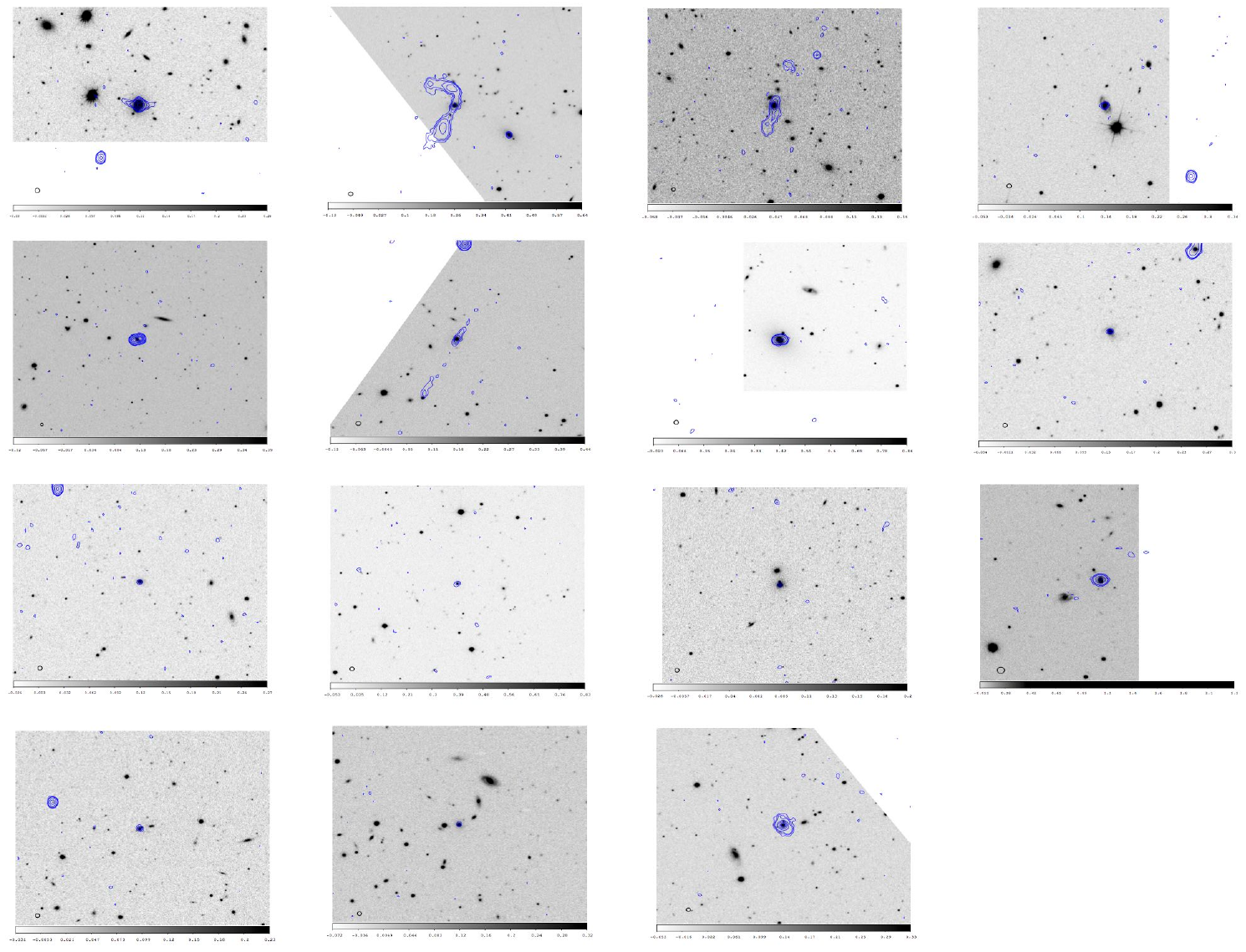}
	\caption{The radio contours overlaid on SDSS r-band image for evolving or un-relaxed (blue) groups. Contour levels of 3, 5, 7, 19, 40, 80, 120, 180, 300, 500, 800, 1000 ${\sigma}$ are shown. The \textit{rms} noise ranges from 0.1-0.2 mJy in sample.}
	\label{Un_relaxed:fig}
\end{figure*}

For comparison of our observational results with the model prediction,  we select dynamically relaxed and un-relaxed galaxy groups in our semi-analytic model on the bases of their mass assembly. According to \cite{Raouf2014}, the groups are classified as dynamically relaxed (old) and un-relaxed (young), when the halos accumulated $>50$ per cent and  $< 30$ per cent of their final mass at  z $\sim$ 1, respectively.  Figure \ref{fig6} shows, the BGG peak and integrated radio luminosity (at 1.4 GHz) as a function of their stellar mass, in analogy to Figure \ref{fig2}. Both the observed data and the model predictions are shown. For the calibration of the model luminosity we used \cite{Best2012} catalog of radio galaxies which extends to z=0.7.

Given that the jet power in radio galaxies is correlated directly with the accretion rate of the super-massive blackhole, the consistently between observations and model predictions suggest a higher rate of accretion for the central blackhole hosted by the most massive galaxy in dynamically evolving (young) galaxy groups relative to those hosted by dynamically evolved (old) galaxy groups at a given stellar mass.

These findings also agree with a recent study of galaxy groups in the Illustris hydrodynamical simulation \citep{Vogelsberger2014a}, in which we found a lower rate of blackhole accretion for a given stellar mass of the BGGs in comparison to the BGGs in young galaxy groups \citep{Raouf2016a}.  

\section{Discussion and conclusions}

Using a sample of galaxy groups from the GAMA survey we demonstrate that the radio luminosity of the most luminous galaxies, usually found in the cores of galaxy groups and clusters, and hence their AGN activities, depends on the dynamical state of the halo. We have used two independent indicators to probe the dynamical state of the halo. The luminosity gap is excepted  to develop as a result of the internal mergers within groups, as argued in the formation of fossil galaxy groups and shown in the cosmological simulations. We found no strong observational support to suggest that the AGN activity in the brightest group galaxies crucially depends on the luminosity gap, alone, evidently developed by major mergers between galaxies in groups. The finding may seen to be in conflict with a recent study of the brightest group galaxies in 610 MHz and 1.4 GHz which point at a similar difference in the radio luminosity of the BGGs in fossil and other galaxy groups \citep{Miraghaei2014}. However its important to note that the sample in the later study satisfies the large luminosity gap criterion, explicitly, but it also satisfies the small BGG offset criterion, implicitly, because the BGGs are located at the peak of the X-ray emission. Therefore the small sample studied by \citep{Miraghaei2014} can be classified at dynamically relaxed groups in an analogy to this study. A more recent study of the IGM properties in fossil groups based on {\em Chandra} X-ray observations shows that the majority of the fossils harbor weak cool-cores \citep{Bharadwaj15}. This confirms our earlier findings on the IGM temperature within the core of the fossil groups \citep{Khosroshahi2004, Khosroshahi2007} and the absence of strong cool-cores in fossils. This study rules out current/strong AGN activities in fossil group dominant galaxies.

The most plausible explanation for our results is that while recent major mergers could be one of the driving phenomena behind the reduced AGN fueling, however the dynamical state of the group, e.g. the combination of the large luminosity gap and the virialization of the halo, is the key driver behind the observed lack of AGN activity probed by the radio emission. Tracing the evolution of the dynamically relaxed halos in the cosmological simulations \citep{Raouf2014} suggests that their BGGs had their last major merger relatively earlier than the BGGs in the un-relaxed groups. An alternative explanation would be the lack of inherent gas in the BGGs within large luminosity gap groups, however, there is no evidence from the morphological or star formation history of the BGGs to support this argument. The BGGs in relaxed or un-relaxed groups are likely be equally influenced by minor mergers.  

We developed a semi-analytic model for radio AGNs to understand the origin of the observed trend. Our model has been able to reproduce the observed offset in the radio luminosity of the BGGs in dynamically relaxed and un-relaxed groups, to a large extend. Our interpretation of the results is that the super-massive blackhole hosted by the BGGs in dynamically young galaxy groups is subject to a higher rate of accretion, for the same stellar mass budget, in comparison to BGGs in dynamically old galaxy groups.  

We conclude that neither the offset between the position of the brightest group galaxies and the luminosity centroid of the group members, nor the large luminosity gap alone, can be responsible for the large radio luminosity offset between the brightest group galaxies in dynamically relaxed and un-relaxed galaxy groups and they together conspire to conceive the observed offset and thus the observed effect is driven by the difference in both the hot and cold accretion modes. 

\acknowledgments
We thank the anonymous referee for his/her constructive comments and suggestions which helped to improve the manuscript. GAMA is a joint European-Australasian project based around a spectroscopic campaign using the Anglo-Australian Telescope. The GAMA input catalog is based on data taken from the Sloan Digital Sky Survey and the UKIRT Infrared Deep Sky Survey. Complementary imaging of the GAMA regions is being obtained by a number of independent survey programs including GALEX MIS, VST KiDS, VISTA VIKING, WISE, Herschel-ATLAS, GMRT and ASKAP providing UV to radio coverage. GAMA is funded by the STFC (UK), the ARC (Australia), the AAO, and the participating institutions. The GAMA website is http://www.gama-survey.org/. The Radio Semi-Analytic Galaxy Evolution (Radio-SAGE) model used in this study is a publicly available for download at https://github.com/mojtabaraouf/sage. The Millennium Simulation was carried out by the Virgo Supercomputing Consortium at the Computing Centre of the Max Plank Society in Garching.

\end{document}